\documentclass[aps,prd,12pt,citeautoscript,superscriptaddress,showpacs,scrartcl]{revtex4-2}

\usepackage{amssymb,latexsym,amsmath,enumerate,amsthm}
\usepackage{graphicx}
\usepackage{color}
\usepackage[colorlinks=true, pdfstartview=FitV,linkcolor=blue, citecolor=blue, urlcolor=blue]{hyperref}

\newcommand{\be}{\begin{equation}}
\newcommand{\ee}{\end{equation}}
\newcommand{\bea}{\begin{eqnarray}}
\newcommand{\eea}{\end{eqnarray}}
\newcommand{\non}{\nonumber}
\newcommand{\Non}{\nonumber\\}
\newcommand{\nn}{\nonumber}

\newcommand{\red}{\textcolor{black}}

\begin{document}


\title{Anisotropic cosmological models in Horndeski gravity}

\author{Rafkat Galeev}
\email{rafgaleev3@gmail.com}
\affiliation{Department of Physics, Kazan Federal University, Kremlevskaya str. 18, Kazan 420008, Russia}

\author{Ruslan Muharlyamov}
\email{rmukhar@mail.ru}
\affiliation{Department of Physics, Kazan Federal University, Kremlevskaya str. 18, Kazan 420008, Russia}

\author{Alexei~ A.~Starobinsky}
\email{alstar@landau.ac.ru}
\affiliation{L.D. Landau Institute for Theoretical Physics RAS, Moscow 119334, Russia}
\affiliation{Department of Physics, Kazan Federal University, Kremlevskaya str. 18, Kazan 420008, Russia}

\author{Sergey V. Sushkov}
\email{sergey$_\,$sushkov@mail.ru}	
\affiliation{Department of Physics, Kazan Federal University, Kremlevskaya str. 18, Kazan 420008, Russia}

\author{Mikhail~S.~Volkov}
\email{volkov@lmpt.univ-tours.fr}
\affiliation{Institut Denis Poisson, UMR - CNRS 7013, \\ Universit\'{e} de Tours, Parc de Grandmont, 37200 Tours, France}
\affiliation{Department of Physics, Kazan Federal University, Kremlevskaya str. 18, Kazan 420008, Russia}


\begin{abstract}
	
It was found recently that the anisotropies in the homogeneous Bianchi~I cosmology considered within the context of a specific Horndeski theory 
are damped  near the initial singularity instead of being amplified. In this work we 
extend the analysis of this phenomenon to cover the whole of the  Horndeski family. 
We find that the phenomenon  is absent in the K-essence and/or Kinetic Gravity Braiding theories, where 
 the anisotropies grow as one  approaches the singularity. The anisotropies are damped 
at early times only in more general Horndeski models whose Lagrangian includes terms 
 quadratic and cubic in second derivatives of the scalar field. Such theories are often considered as being 
 inconsistent with the observations because they predict a non-constant speed of gravitational waves. 
 However, the predicted value of the speed {\it at present} can be
 close to  the speed of light 
 with any required precision, hence the theories  actually  agree with the present time observations. 
 We consider two different examples of such theories, both characterized by 
 a late  self-acceleration and an early inflation  driven by the non-minimal coupling. 
 Their anisotropies are maximal  at intermediate times and approach zero 
 at early and late times.  The early inflationary stage exhibits  an instability with respect to 
 inhomogeneous perturbations, suggesting that  the initial  state of the universe should be inhomogeneous. 
However,  more general  Horndeski models may probably be stable. 
\end{abstract}

\maketitle

\section{Introduction} 
\setcounter{equation}{0}

It is usually assumed that the state of the universe close to the initial singularity should be strongly 
anisotropic \cite{Belinsky:1970ew,Collins:1972tf,Belinsky:1982pk}. 
This belief  is based on the fact that spatial anisotropies produce in 
the Einstein equations terms  
which become dominant when one goes backwards in time. In other words, anisotropic perturbations 
grow to the past.  When the universe expands, the anisotropy  terms decrease faster  than  the 
contribution of other forms of energy subject to the dominant energy condition, and the universe rapidly approaches 
a locally  isotropic state during inflation \cite{Starobinsky:1982mr}, \cite{Wald:1983ky}
(without the inflationary stage this process may require a longtime or may not happen at all due to the 
possibility of recollapse). 
Therefore, thinking about the early  history of the universe, one could expect 
the isotropic phase of inflation to be generically  preceded by an anisotropic 
phase. 

Although this argument seems quite robust, an explicit example in which 
the anisotropies in the Bianchi I homogeneous model  are  damped at early times instead of being amplified 
was recently found \cite{Starobinsky:2019xdp} within the context of a 
specific Horndeski theory for a gravitating scalar field \cite{Horndeski:1974wa}. 
Therefore, the initial stage of the universe in this theory is not anisotropic. 

It remained unclear whether the finding of \cite{Starobinsky:2019xdp} is generic or  specific only for one 
particular Horndeski model. To find the answer,  we extend in what follows the analysis of \cite{Starobinsky:2019xdp} 
to cover the whole of the  Horndeski family. 
We find that the effect of the anisotropy damping is not necessarily present in all Horndeski theories. In particular, it is 
absent in the the K-essence and/or Kinetic Gravity Braiding theories. 
The spatial  anisotropies in such theories always grow as one  approaches the singularity. 
 However, the anisotropies are damped 
at early (and late) times  in more general Horndeski models whose Lagrangian includes terms 
 quadratic and cubic in second derivatives of the scalar field. Such theories are often considered as being 
 inconsistent with the observations because they predict 
 a non-constant speed of gravitational waves (GW) \cite{Creminelli:2017sry,Ezquiaga:2017ekz,Baker:2017hug}, 
whereas the  GW170817 event shows that the GW speed is equal to the speed of light with very high precision \cite{GW}. 
 However, the theories actually predict the value of the GW speed {\it at present} to be close to unity 
 within the required precision. In addition, the theories admit stable in the future self-accelerating cosmologies. 
 Therefore, they can perfectly agree with the current observations, 
and we can extrapolate them to the early times as well since no observational data  about the GW speed 
at redshifts $z>0.3$ are currently available. 

We consider two different examples of such theories, both characterized by a late time self-acceleration and also 
by an  early time inflation driven by the nonminimal couplings
arising in the Horndeski theory. Sometimes this phase is called ``kinetic inflation" \cite{Sushkov:2009hk}. 
The anisotropies in these theories show a maximum at intermediate times and approach zero 
at early and late times. Therefore, the early universe cannot be anisotropic, but it 
cannot be isotropic either  since it  is unstable with respect to the inhomogeneous perturbations.
This suggests that the initial phase should be inhomogeneous. At the same time, it remains unclear if the gradient 
instabilities at early times are omnipresent in all Horndeski models. 
One of the two models that we consider has less instabilities than the other,
therefore, it is conceivable that  some other more general Horndeski theories may be completely stable.


\section{Horndeski theory}
\setcounter{equation}{0}

This is the most general theory for a  gravity-coupled scalar field $\phi$ whose equations are at most of second order. 
The theory was first obtained in \cite{Horndeski:1974wa}, but we 
shall use its action in the form given in  \cite{Kobayashi:2011nu}: 
\be
\mathcal{S}=\int{}\;( {\cal L}_2+{\cal L}_3+{\cal L}_4+{\cal L}_5)\, \sqrt{-g}\,d^4x\,,
\label{action}
\ee
where 
\bea
{\cal L}_2 &=& G_2(\phi,X)\,,
\non\\
{\cal L}_3 &=& - G_3(\phi,X)\Box\phi\,,
\non\\
{\cal L}_4 &=& G_{4}(\phi,X) R +G_{4X}(\phi,X)
\left[ (\square \phi )^{2}-(\nabla_\mu \nabla_\nu \phi)^2  \right],
\non\\
{\cal L}_5 &=& G_{5}(\phi,X) G_{\mu\nu}\,\nabla^\mu \nabla^\nu \phi -\frac{1}{6}G_{5X}\times
\Non
&& \times \left[\left( \Box \phi \right)^3 -3 \Box \phi (\nabla_\mu \nabla_\nu \phi)^2 + 2\left(\nabla_\mu \nabla_\nu \phi \right)^3 \right]. 
\label{L_i}
\eea
Depending on the choice of the four 
arbitrary functions $G_{A}(\phi,X) $ (with $A=2,3,4,5$)  of the scalar field $\phi$ and of its canonical kinetic term
$X=-\frac{1}{2}\nabla^\mu\phi \nabla_\mu\phi$,
this determines not just one theory but a large family of theories. 
One has  $G_{AX}\equiv \partial G_A/\partial X$, $(\nabla_\mu \nabla_\nu \phi)^2=\nabla_\mu \nabla_\nu \phi  \,\nabla^\nu \nabla^\mu \phi$, and $\left(\nabla_\mu \nabla_\nu \phi \right)^3= \nabla_\mu \nabla_\nu \phi  \,\nabla^\nu \nabla^\rho \phi \, \nabla_\rho \phi  \nabla^\mu \phi$. Finally, $R$ and $G_{\mu\nu}$ are the Ricci scalar  and the Einstein tensor.  

For example, setting $G_3=G_5=0$, $G_4=const$ and $G_2=X-V(\phi)$ yields the standard theory of the inflaton type, 
a more general choice of $G_2(\phi,X)$ yields the $K$-essence theory \cite{ArmendarizPicon:1999rj}, 
while including also $G_3(\phi,X)\neq 0$ yields the Kinetic Gravity Braiding (KGB) theory \cite{Deffayet:2010qz}. 
The KGB theory, possible with $G_4=G_4(\phi)$, is the most general Horndeski model in which the sound speed of 
tensor perturbations is equal to the speed of light \cite{Creminelli:2017sry,Ezquiaga:2017ekz,Baker:2017hug}. 
The Lagrangian of this theory contains the second derivatives of the scalar field only linearly.  
If the Lagrangian contains also quadratic $(\nabla_\mu \nabla_\nu \phi)^2$ and/or cubic $\left(\nabla_\mu \nabla_\nu \phi \right)^3$ terms,
which is the case if $G_4$ and/or $G_5$ depend on $X$, then the GW speed  is no longer constant.

\section{Bianchi I model}
\setcounter{equation}{0}

The simplest cosmological model is homogeneous and isotropic, with the metric 
\be
ds^2=-{\rm N^2}\, dt^2+{\rm a}^2\,(dx_1^2+dx_2^2+dx_3^2),
\ee
where the scale factor ${\rm a}$, the lapse ${\rm N}$, as well as the scalar field $\phi$, depend only on $t$.  
The corresponding field equations for the theory \eqref{action} are explicitly shown in \cite{Kobayashi:2011nu}. 
We make the next step and consider the homogeneous and anisotropic 
 Bianchi I metric,
\be\label{ma}
ds^2=-{\rm N^2}\,dt^2+{\rm a}^2_1\,\,dx_1^2+{\rm a}^2_2\,\,dx_2^2+{\rm a}^2_3\,\,dx_3^2\,,
\ee
with the three scale factors ${\rm a}_m$ ($m=1,2,3$), the lapse ${\rm N}$ and the scalar field $\phi$ depending only on $t$.  
Substituting this into \eqref{action} yields the reduced one-dimensional action that can be varied with respect to ${\rm a}_m$, 
${\rm N}$ and  $\phi$. Although the action contains the second derivatives,  
all higher derivatives arising during the variation cancel. As a result, first varying the action with respect to ${\rm N}$ 
and ${\rm a}_m$ 
and then imposing the gauge condition ${\rm N}=1$, yields the following equations:
\begin{eqnarray}
\label{00}
G^0_0\left({\cal G}-2G_{4X}\dot{\phi}^2 - 2G_{4XX}\dot{\phi}^4 
+2G_{5\phi}\dot{\phi}^2 +G_{5X\phi}\dot{\phi}^4\right) = 
G_2 - G_{2X}\dot{\phi}^2
\Non
 - 3G_{3X}H\dot{\phi}^3 + G_{3\phi}\dot{\phi}^2 
  + 6G_{4\phi}H\dot{\phi} + 6G_{4X\phi}\dot{\phi}^3 H && 
\Non  
 - 5G_{5X}H_1H_2H_3\dot{\phi}^3 - G_{5XX}H_1H_2H_3\dot{\phi}^5, 
&&
\\
\label{ii}
{\cal G} G^{i}_{i}-(H_{j}+H_{k})\frac{d{\cal G}}{dt} =
G_2 - \dot\phi \frac{d G_3}{dt} 
+2 \frac{d}{dt}(G_{4\phi}\dot{\phi}) 
-\frac{d}{dt}(G_{5X}\dot{\phi}^3 H_{j}H_{k}) 
&&
\Non
-G_{5X}\dot{\phi}^3 H_{j}H_{k}(H_{j}+H_{k})
\,.&&
\end{eqnarray}
Here the dot denotes  the $t$-derivative, one has $H_i=\dot {\rm a}_i/{\rm a}_i$, and 
the average Hubble parameter is $H=\frac{1}{3}\sum_{i=1}^3 H_i\equiv \dot{\rm a}/{\rm a}$  
with  ${\rm a}=({\rm a}_1{\rm a}_2{\rm a}_3)^{1/3}$. The Einstein tensor components are
\begin{eqnarray}
&&G^0_0=-\left(H_1H_2 +H_2H_3 +H_3H_1\right)\,,
\Non
&&G^{i}_{i}=-\left(\dot{H}_{j} +\dot{H}_{k} +H_{j}^2 +H_{k}^2 +H_{j}H_{k}\right)\,,
\end{eqnarray}
where the triples of indices $\{i,j,k\}$ take values $\{1,2,3\}$, $\{2,3,1\}$, or $\{3,1,2\}$.
In addition, we have defined 
\begin{equation}\label{def_G}
{\cal G} = 2G_4-2G_{4X}\dot{\phi}^2+G_{5\phi}\dot{\phi}^2\,.
\end{equation}

Varying the action \eqref{action} with respect to $\phi$ yields the equation which,
after some rearrangements,  can be cast into the following form:
\begin{eqnarray}\label{scalar}
\frac{1}{\rm a^3}\frac{d}{dt}({\rm a^3} {\cal J})={\cal P}\,,
\end{eqnarray}
with 
\begin{eqnarray}
\label{J}
{\cal J} &=&
\dot{\phi}\, \Big[ G_{2X}-2G_{3\phi} +3H\dot\phi(G_{3X} -2G_{4X\phi})
\Non
&&
+G^0_0(-2G_{4X}-2\dot\phi^2G_{4XX} +2G_{5\phi} +G_{5X\phi}\dot\phi^2 )
\notag\\
&& +H_1H_2H_3(3G_{5X}\dot{\phi} +G_{5XX}\dot{\phi}^3)\Big]\,,
\\
\label{P}
{\cal P} &=& G_{2\phi} -\dot{\phi}^2(G_{3\phi\phi} 
+G_{3X\phi}\ddot \phi)
+RG_{4\phi}+2G_{4X\phi}\dot\phi(3\ddot\phi H-\dot\phi
G^0_0)
\Non
&&+G^0_0G_{5\phi\phi}\dot{\phi}^2+G_{5X\phi}\dot{\phi}^3H_1H_2H_3\,, 
\end{eqnarray}
where $R$ is the scalar curvature, {$R=-G_\mu^{\mu}$.}

Let us parameterize  the three scale factors as
\begin{equation}\label{factorParam}  
{\rm a}_1 = {\rm a}\, e^{\beta_{+}+\sqrt{3}\beta_{-}},\quad 
{\rm a}_2 = {\rm a}\, e^{\beta_{+}-\sqrt{3}\beta_{-}},\quad 
{\rm a}_3 = {\rm a}\, e^{-2\beta_{+}}\,,
\end{equation}
hence  
\begin{eqnarray}\label{HubbleParam}
H_1=H+\dot{\beta}_{+}+\sqrt{3}\dot{\beta}_{-}\,,\,~~~
H_2=H+\dot{\beta}_{+}-\sqrt{3}\dot{\beta}_{-}\,, \,~~~
H_3=H-2\dot{\beta}_{+}\,,
\end{eqnarray}
where $H=\dot{\rm a}/{\rm a}$.  The anisotropies are determined by $\dot{\beta}_\pm$, and if they vanish, then 
$H_1=H_2=H_3=H$ and the universe is isotropic. 
It will be convenient to introduce 
\begin{equation}\label{sigma} 
\sigma^2 = \dot{\beta}^2_{+} + \dot{\beta}^2_{-}\,.
\end{equation}
Using these definitions, the $G^0_0$ Einstein equation \eqref{00}
assumes the form 
\bea       \label{eq1}
3\big(H^2-\sigma^2\big)
\left({\cal G}-2G_{4X}\dot{\phi}^2-2G_{4XX}\dot{\phi}^4
+2G_{5\phi}\dot{\phi}^2
\red{+G_{5X\phi}\dot{\phi}^4 }
\right) =-G_2 +\dot{\phi}^2G_{2X}
&&
\Non
 +3G_{3X}H\dot{\phi}^3 -G_{3\phi}\dot{\phi}^2 -6G_{4\phi}H\dot{\phi} -6G_{4X\phi}H\dot{\phi}^3
&&
\Non
+\dot{\phi}^3(5G_{5X} \red{+}G_{5XX}\dot{\phi}^2)
(H - 2\dot{\beta}_{+})
\big[(H+\dot{\beta}_+)^2 -3\dot{\beta}_{-}^2\big]\,.
&&
\eea
This equation contains only first derivatives. 
The remaining three Einstein equations \eqref{ii} contain second derivatives and read 
\bea
\big(2\dot{H}+3H^2 + 3\sigma^2\big){\cal G}+2H\dot{\cal G}
=
-G_2+G_{3\phi}\dot{\phi}^2 +G_{3X}\dot{\phi}^2\ddot{\phi} -2\frac{d}{dt}\big(G_{4\phi}\dot{\phi}\big)
&&
\Non
\red{+\frac{d}{dt}\left[G_{5X}\dot{\phi}^3 \left(H^2-\sigma^2\right)\right]}
\red{+2G_{5X}\dot{\phi}^3\left(H^3+\dot{\beta}_{+}^3-3\dot{\beta}_{+}\dot{\beta}_{-}^2 \right)},  \label{eq2}
&&
\\
\label{eq_bp}
\frac{d}{dt}\left[{\cal G}{\rm a}^3\dot{\beta}_{+} 
+G_{5X}\dot\phi^3 {\rm a}^3 \left(\dot\beta_{-}^2-\dot\beta_{+}^2 -H\dot\beta_{+}\right)  
\right]=0,
&&
\\
\label{eq_bm} 
\frac{d}{dt}\left[{\cal G}{\rm a}^3\dot{\beta}_{-} 
+G_{5X}\dot\phi^3 {\rm a}^3 \left(2\dot\beta_{+}\dot\beta_{-} -H\dot\beta_{-}\right)  
\right]=0.
&&
\eea
We notice that the two latter equations have the total derivative structure and can be integrated once, which gives 
first order conditions 
\bea
\label{fi1}
{\cal G}\dot{\beta}_{+} 
+G_{5X}\dot\phi^3  \left(\dot\beta_{-}^2-\dot\beta_{+}^2 -H\dot\beta_{+}\right) &=& \frac{C_+}{{\rm a}^3}\,,
\\
\label{fi2}
{\cal G}\dot{\beta}_{-} 
+G_{5X}\dot\phi^3  \left(2\dot\beta_{+}\dot\beta_{-} 
-H\dot\beta_{-}\right) &=& \frac{C_-}{{\rm a}^3}\,,
\eea
with $C_+,C_-$ being  integration constants. Supplementing 
these two equations by the first order equation \eqref{eq1} 
and by the scalar field equation \eqref{scalar}, yields a closed system of four differential equations 
for the four functions ${\rm a}(t),\beta_\pm(t)$ and $\phi(t)$. The remaining equation \eqref{eq2}
can be ignored, since it is  automatically fulfilled by virtue of the Bianchi identities.

An additional simplification is achieved if  the  scalar source ${\cal P}$ defined by \eqref{P} vanishes, 
since in this case the scalar field equation \eqref{scalar}
also  assumes the total derivative structure and can be integrated once. The source ${\cal P}$ will vanish if 
all four functions $G_A$ are independent on $\phi$, in which case the theory 
is invariant under shifts 
$
\phi\to \phi+\phi_0. 
$
However, ${\cal P}$ will vanish also if $G_2$ and $G_4$ are independent of $\phi$, 
while  $G_3$ and $G_5$ depend on $\phi$ only {\it linearly},  such that $G_{3\phi}=const$  and  $G_{5\phi}=const$.
Then the scalar field equation \eqref{scalar} becomes 
\bea         \label{fi3}
\dot{\phi}\, \Big[ G_{2X}+3HG_{3X} \dot\phi 
+G^0_0(-2G_{4X}-2\dot\phi^2G_{4XX} +2G_{5\phi} ) &&
\notag\\
+(H-2\dot\beta_+)[(H+\dot\beta_+)^2-3\dot\beta_-^2]
(3G_{5X}\dot{\phi} +G_{5XX}\dot{\phi}^3)\Big]+\frac{C_\phi}{{\rm a}^3}&=&0,
\eea 
with $C_\phi$ being an integration constant. 
The problem therefore reduces in this case to four equations \eqref{eq1},\eqref{fi1},\eqref{fi2} and \eqref{fi3} 
which determine {\it algebraically}  the Hubble parameter $H({\rm a})$, the anisotropies $\dot\beta_\pm({\rm a})$, 
and the derivative of the scalar field $\dot\phi({\rm a})$.

To recapitulate, if there is an explicit dependence on $\phi$, then the problem reduces to four 
differential equations \eqref{eq1},\eqref{fi1},\eqref{fi2} and 
\eqref{scalar} to determine ${\rm a}(t)$, $\beta_\pm(t)$, $\phi(t)$. 
If the coefficient functions $G_2,G_4$ are $\phi$-independent while  $G_3,G_5$ depend on $\phi$ at most
linearly, then the problem reduces to four equations  \eqref{eq1},\eqref{fi1},\eqref{fi2},\eqref{fi3} 
which determine 
the functions $H({\rm a})$, $\dot\beta_\pm({\rm a})$ and $\dot\phi({\rm a})$ {\it algebraically}. The time dependence 
can then be restored by integrating the equation $\dot{\rm a}/{\rm a}=H({\rm a})$. 

In what follows we shall not at first assume anything about the $\phi$-dependence, but later we shall consider
specific examples admitting the simplified description in terms of the four algebraic equations. 
Our aim is to study the anisotropies described by \eqref{fi1} and \eqref{fi2}. The structure of these equations 
suggests considering separately two different  cases, $G_{5X}=0$  and $G_{5X}\neq 0$, 
which will be described, respectively, in the following two Sections. 
 

\section{The $G_{5X}=0$ case}
\setcounter{equation}{0}

In this case the anisotropy equations  \eqref{fi1} and \eqref{fi2} are linear in $\dot\beta_\pm$ and yield 
\be                \label{bet}
\dot{\beta}_{\pm} = \frac{C_\pm}{{\cal G}\,{\rm a}^3}.
\ee
The behaviour of the anisotropies  is therefore determined by the function ${\cal G}$ defined by \eqref{def_G}. This definition can 
equivalently be viewed as the equation for $G_4$, 
\be        \label{ee}
{\cal G}(\phi,X)=2G_4(\phi,X)-4X\frac{\partial G_4(\phi,X)}{\partial X}+2XG_{5\phi}(\phi), 
\ee
whose solution is 
\be           \label{ee1}
G_4(\phi,X)=f(\phi)\sqrt{X}+g^\prime(\phi)\,X-\frac{\sqrt{X}}{4}\int \frac{{\cal G}(\phi,X)}{X^{3/2}}\,dX,~~~~~G_5=g(\phi), 
\ee
with arbitrary $f(\phi)$ and $g(\phi)$. 
Let us first consider the subcase where

\subsection{${\cal G}=\mu=const$}
In this case Eq.\eqref{bet} yields 
\be                \label{bet1}
\dot{\beta}_{\pm} = \frac{C_\pm}{\mu \,{\rm a}^3},
\ee
so that the anisotropies behave in the same way as in General Relativity: they grow as ${\rm a}\to 0$. 
Therefore, the initial singularity is strongly anisotropic, while at late times the anisotropies decay. 
Eq.\eqref{ee} then yields 
\be           \label{ee2}
G_4(X)=\frac{\mu}{2}+f(\phi)\sqrt{X}+g^\prime(\phi)\,X,~~~~~~G_5=g(\phi). 
\ee
This describes all the conventional theories. Setting $f(\phi)=g(\phi)=0$ one can, depending on whether 
$G_2$ and $G_3$ are included or not, distinguish the following particular cases.
\begin{itemize}
\item 
$G_2=G_3=G_5=0$, $G_4=\mu/2$. This corresponds to the vacuum General Relativity, assuming that $\mu=M_{\rm Pl}^2$. 
\item 
$G_2=X-V(\phi)$ and $G_3=G_5=0$, $G_4=\mu/2$, which   defines  the General Relativity with the conventional scalar field. 
\item 
$G_2(\phi,X)$ and $G_3=G_5=0$, $G_4=\mu/2$, which gives   the K-essence theory. 
\item 
$G_2(\phi,X)$,  $G_3(\phi,X)$, $G_5=0$, $G_4=\mu/2$, which  gives  the KGB theory. 
\end{itemize}
In all of these theories the anisotropies $\dot\beta_\pm$ grow  as one approches the initial singularity.

\subsection{${\cal G}=\mu(\phi) $}
Formulas \eqref{bet},\eqref{ee2} still apply, with the  replacement $\mu\to\mu(\phi)$. 
Let us consider the simplest option:
\be
G_2=X,~~~G_3=G_5=0,~~~~G_4=\frac12\,\mu(\phi),~~~~\dot{\beta}_{\pm} = \frac{C_\pm}{\mu(\phi) \,{\rm a}^3}.
\ee
Since $G_4$ depends on $\phi$, the $\phi$-equation remains differential and 
the system does not reduce to algebraic equations. 
At the same time, the theory with the gravitational kinetic term $\mu(\phi) R$  can be converted to the theory 
with the standard kinetic term $\mu R$  by a conformal transformation of the metric. This brings  us back to the theories 
considered in the previous subsection, where the anisotropies are always unbounded near singularity. 
Performing the inverse conformal transformation to pass to the original frame changes only the scale  factor 
(and the proper time) without changing the anisotropies. Hence the latter are unbounded in the original frame too. 
Therefore, the choice  ${\cal G}=\mu(\phi)$
does not insure the damping of anisotropies, and we shall now consider a more complex choice.

\subsection{${\cal G}={\cal G}(X)$ and $G_4(X)$}

We shall consider the theory sometimes called ``kinetic inflation"
\cite{Sushkov:2009hk,Sushkov:2012,Saridakis:2010mf,Skugoreva:2013ooa}, 
\cite{Gao:2010vr,Granda:2010hb,Sadjadi,Banijamali,Gubitosi:2011sg}. 
It corresponds   to the choice 
\be              \label{XXX}
G_2=X-\Lambda,~~~~~~G_3=0,~~~~~G_4=\frac12\left(\mu+\gamma X\right),~~~~~~G_5=\frac12\,(\alpha+\gamma)\,\phi,
\ee
where $\Lambda$, $\mu,\alpha$ are constant parameters. 
The constant  $\gamma$ is a  gauge parameter
which  drops out from the equations due to 
the relation 
$
XR+(\Box\phi)^2-(\nabla_\mu\nabla_\nu\phi)^2=-\phi G_{\mu\nu}\nabla^\mu\nabla^\nu\phi+\text{\it total derivative} 
$
\cite{Kobayashi:2011nu}, which allows one to trade the $G_5\sim\phi$ term in the Lagrangian \eqref{L_i} for the $G_4\sim X$ term. 
In the $\gamma=0$ gauge one has  $G_4=const$ and $G_5\sim\phi$, 
while choosing $\gamma=-\alpha$ yields $G_5=0$.

The homogeneous and isotropic cosmologies in the model \eqref{XXX} are characterized,  apart from the 
late inflationary phase driven by $\Lambda$,  also  by an early  inflationary phase with the Hubble rate 
determined not by $\Lambda$ but rather by $\alpha$, so that $\Lambda$ is ``screened at early times"
\cite{Starobinsky:2016kua}.
The GW speed in the theory is not constant, but its value at present is predicted to be 
 close to the speed of light with a very high precision
\cite{Starobinsky:2019xdp}.

Injecting \eqref{XXX} to \eqref{ee} yields 
\be
{\cal G}=\mu+\alpha X~~~~~~\Rightarrow~~~~~\dot\beta_\pm=\frac{C_\pm}{(\mu+\alpha X){\rm a}^3}.
\ee
It turns out that   $X=\dot{\phi}^2/2$  grows fast enough for ${\rm a}\to 0$ to suppress the anisotropies \cite{Starobinsky:2019xdp}.

Let us write down explicitly what becomes to the equations 
\eqref{eq1},\eqref{fi1}-\eqref{fi3}: 
\bea
3(H^2-\dot{\beta}_{+}^2-\dot{\beta}_{+}^2)\left(\mu+\frac32\,\alpha\, \dot{\phi}^2\right)&=&\frac{1}{2}\,\dot{\phi}^2+\Lambda\,,\nn \\
\left(3\alpha\, (H^2-\dot{\beta}_{+}^2-\dot{\beta}_{+}^2)-1\right)\dot{\phi}&=&\frac{C_\phi}{{\rm a}^3}\,,\nn \\
\left(\mu+\frac{\alpha}{2}\,\dot{\phi}^2\right)\dot{\beta}_\pm &=&\frac{C_\pm}{{{\rm a}^3}}\,.          \label{EEE}
\eea 
We shall need a dimensionless version of these equations. Let us assume that $\alpha>0$.
If 
$H_0$ and ${\rm a}_0$ are the present values of the Hubble parameter and of the scale factor, then setting 
\bea
{\rm a}={\rm a}_0\,a,~~~H=H_0\,\sqrt{y},~~~~\Lambda=3\mu H_0^2\Omega_0,~~~~~~
\alpha=\frac{1}{3H_0^2\zeta},
\nn \\
C_\phi=\sqrt{18\mu\alpha \Omega_6} H_0^2 {\rm a}_0^3,~~~~C_\pm =\mu H_0 {\rm a}_0^3 Q_\pm\,,~~~~
\dot{\phi}=\sqrt{\frac{2\mu\Omega_6}{\alpha}}\,\psi\,,~~~~\dot{\beta}_\pm=H_0 s_\pm
\eea
reduces \eqref{EEE} to equations containing only dimensionless variables $a,Y,\psi,s_\pm$ and dimensionless parameters 
$\zeta,\Omega_0,\Omega_6$:
\bea
\Omega_6 \left(3Y-\zeta\right)\psi^2+Y=\Omega_0\,,~~~
(Y-\zeta)\psi&=&\frac{1}{a^3},~~~~~
(\Omega_6 \psi^2+1)s_\pm =\frac{Q_\pm}{a^3}\,,
\eea
where $Y=y-s_{+}^2-s_{-}^2$. 
The solution can be expressed in the parametric form, as functions of $Y$:
\bea                \label{sl}
a^6=\frac{\Omega_6(\zeta-3Y)}{(Y-\zeta)^2 (Y-\Omega_0)},~~~~\psi=\frac{1}{(Y-\zeta)a^3},~~~~
s_\pm= Q_\pm {\cal S},~~~~y=Y+s_+^2+s_-^2\,,~~
\eea
where 
\be             \label{sl1} 
{\cal S}=\frac{1}{(\Omega_6\psi^2+1)a^3}=\frac{(Y-\zeta)^2 a^3}{\Omega_6+(Y-\zeta)^2 a^6}\,.
\ee
When the parameter $Y$ ranges from $\zeta/3$ to $\Omega_0$, the scale factor $a$ changes, respectively, from zero to infinity. 
As one can see, the function ${\cal S}$ determining the anisotropies approaches zero in both of these limits, 
hence the universe becomes isotropic not only at late times but also at early times. In both limits the amplitude $Y$ reduces 
to $y$ and   the Hubble rate is 
\be
\left(\frac{H_{\rm early}}{H_0}\right)^2\equiv \frac{\zeta}{3}\leftarrow y=\left(\frac{H}{H_0}\right)^2\rightarrow \Omega_0\equiv 
\left(\frac{H_{\rm late}}{H_0}\right)^2~~~~\text{as}~~~~~0\leftarrow a\rightarrow \infty. 
\ee
Therefore, the universe interpolates between the early and late isotropic inflationary stages driven  by $\zeta$ and $\Omega_0$,
respectively. The present stage of the universe is highly isotropic, hence $Y\approx y=a=1$ should fulfill \eqref{sl}, 
which requires that 
\be
\Omega_6=\frac{(1-\zeta)^2(1-\Omega_0)   }{(\zeta-3)  }. 
\ee
As a result,  the theory  actually 
depends only on two parameters $\zeta$ and $\Omega_0$ determining values of the two Hubble rates,
apart from the anisotropy charges $Q_\pm$. 
\begin{figure}
     \centering   
       			\resizebox{9cm}{7cm}{\includegraphics{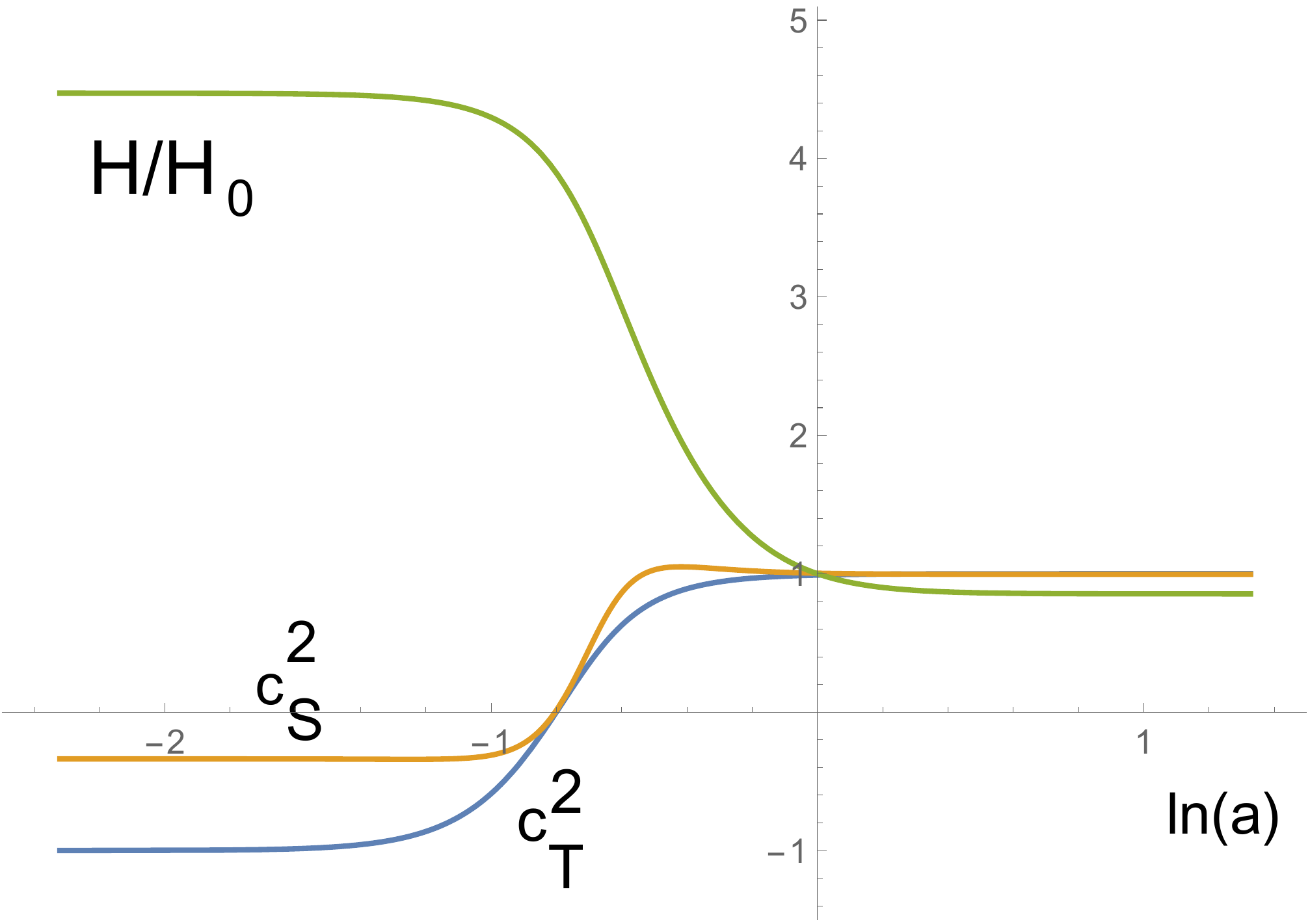}}

   \caption{The dimensionless Hubble rate $\sqrt{y}=H/H_0$, the sound speeds squared in the scalar and tensor sectors 
 against $\ln(a)$ for the isotropic $(Q_\pm=0)$ solution \eqref{sl} with 
   $\Omega_0=0.73$ and $\zeta=60$.}   
   \label{Fig1}
\end{figure}

Setting $Q_\pm=0$ yields homogeneous and isotropic solutions, in which  case one can apply 
the known formulas about describing  small fluctuations. These formulas apply also for anisotropic solutions 
with $Q_\pm\neq 0$ 
at late and early times, 
when the solutions become isotropic. 
The quadratic action for fluctuations is 
\be
I=\frac{\mu}{2}\int {\rm K}\left( \dot{F}^2-c^2\frac{\rm p^2}{{\rm a}^2}\,F^2\right){\rm a}^3 d^4x,
\ee
where $F$ denotes the fluctuation amplitude after separating the variables and ${\rm p}$ is the spatial momentum. 
The expressions for the kinetic term ${\rm K}$ 
and the sound speed squared $c^2$ within the model \eqref{XXX} were derived  in \cite{Starobinsky:2019xdp}, 
and they agree  with  the earlier result obtained 
within the generic Horndeski theory \cite{Kobayashi:2011nu}. It turns out that the kinetic term is always positive,
both in the tensor and scalar sectors, hence there are no ghosts. As seen in Fig.\ref{Fig1}, the sound speeds in both sectors
are not constant, but they  approach
unity at late times. 
 The deviation of the speed of tensor modes from unity at present is negligibly small
 and proportional to $(H_{\rm late}/H_{\rm early})^2$ \cite{Starobinsky:2019xdp}.

It is also worth mentioning that, when written in the gauge where $G_5=0$ and hence  $G_4=(\mu-\alpha X)/2$, the theory \eqref{XXX} 
can be mapped to Class I DHOST  theory \cite{Langlois:2018dxi} 
via a disformal transformation of the metric $g_{\mu\nu}\to \tilde g_{\mu\nu}=A(X)\, g_{\mu\nu}+B(X)\, \nabla_\mu\phi\nabla_\nu\phi$. 
This transformation changes the light cone, hence the sound speeds change. If the functions $A,B$ 
are chosen such that $B(A-2X B)G_4=G_{4X}$ then the resulting DHOST  theory 
will respect the condition 
which insures that the GW speed is equal to the 
speed of light (in the language of \cite{Langlois:2020xbc} this condition is $\alpha_1=\alpha_2=0$; see  Eq.(D.5) of that work). 
Therefore, the GW speed can be made constant via the disformal transformation. 

 \begin{figure}
    
     \centering   
       \includegraphics[scale=0.35]{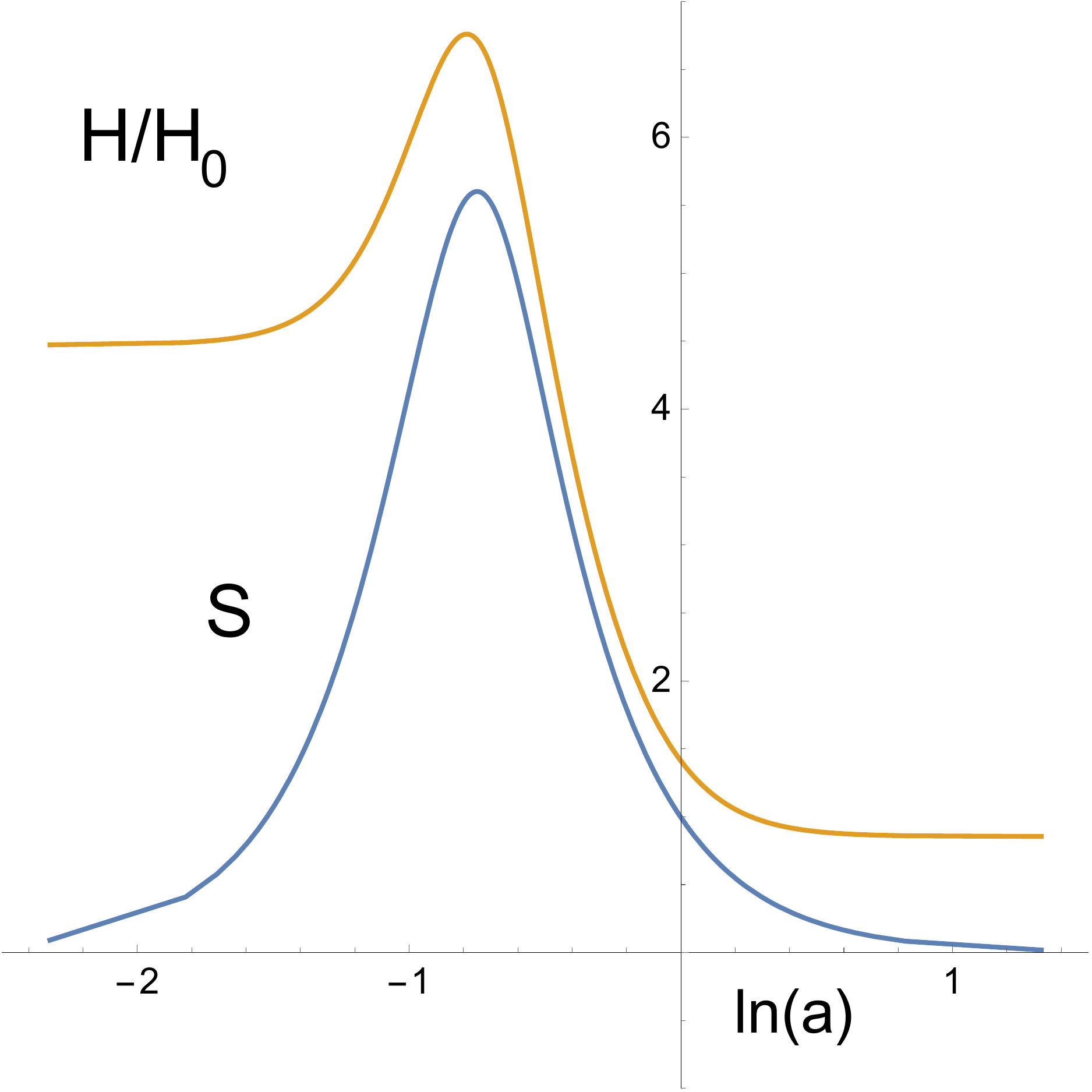}
      \hspace{1 mm}
    \includegraphics[scale=0.35]{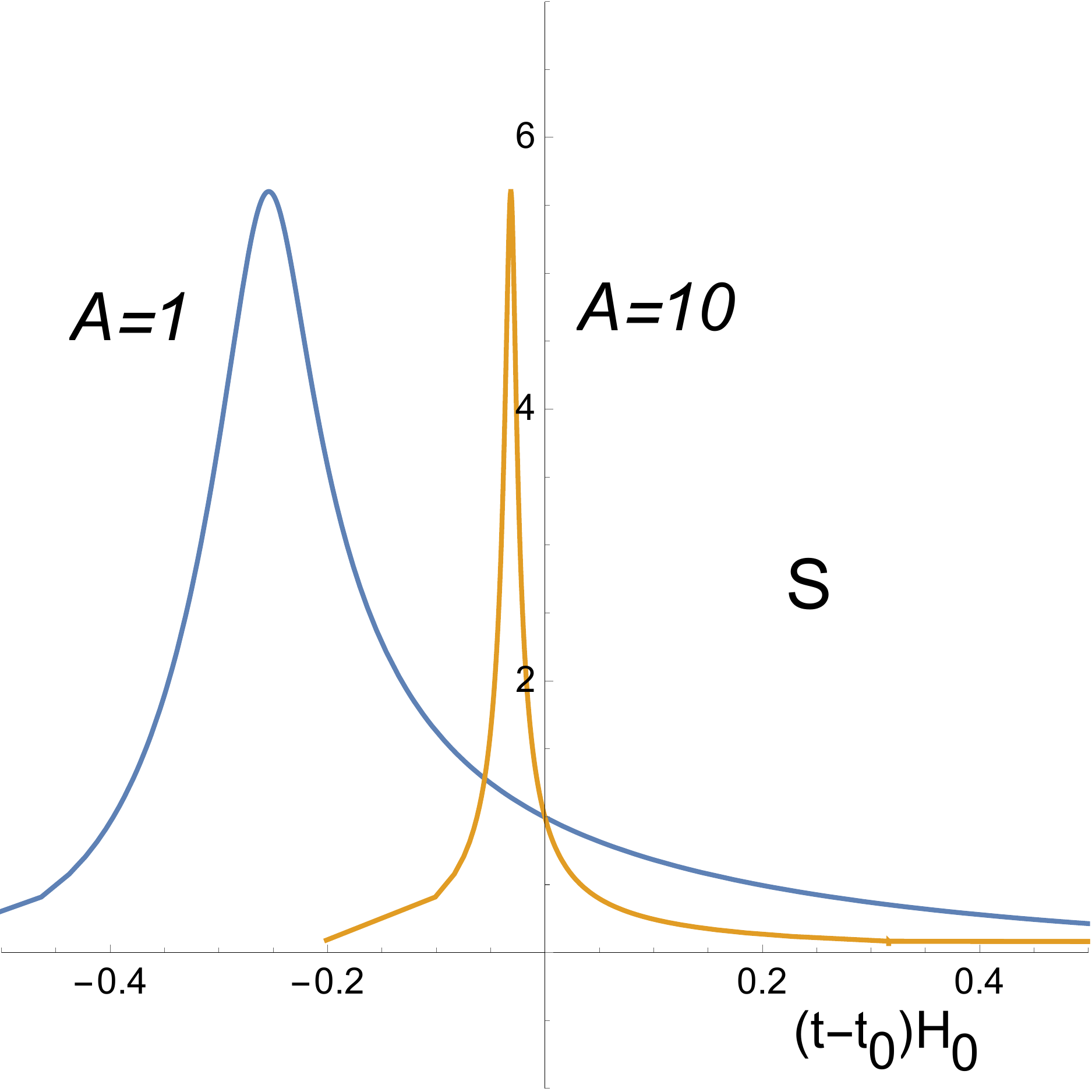}
    
   \caption{Left: the anisotropy amplitude ${\cal S}$  
   and the Hubble rate $H/H_0$  defined by \eqref{sl},\eqref{sl1} 
   with $\Omega_0=0.73$, $\zeta=60$ and ${\cal A}=\sqrt{Q_{+}^2+Q_{-}^2}=1$
   against $\ln(a)$. 
   Right: ${\cal S}(t)$  for ${\cal A}=1$ and ${\cal A}=10$.}   
   \label{Fig2}
\end{figure}

The anisotropies are  $s_\pm =Q_\pm {\cal S}$ where, as seen in Fig.\ref{Fig2},  the function ${\cal S}$ is well localized,
hence the anisotropies vanish both at the early and late stages of the universe and are maximal  in between. 
It is worth noting that, as seen in Fig.\ref{Fig2}, 
the anisotropies contribute to   the Hubble rate and increase it. Since
\be
dt=\frac{1}{H}\,d\ln(a),
\ee
the proper time interval $dt$ decreases if $H$ increases, hence the proper time duration of the anisotropic period 
{\it decreases} when  the anisotropy amplitude ${\cal A}\equiv \sqrt{Q_{+}^2+Q_{-}^2}$ gets larger, since $H$ then increases. 
In other words, the function ${\cal S}(t)$ shows a 
more and more narrow peak when ${\cal A}$ gets larger, as seen in Fig.\ref{Fig2}. 

One should emphasise that, although  the anisotropies approach zero at early times, 
still the universe cannot be isotropic at this stage, 
since it is unstable in this limit with respect to inhomogeneous perturbations. This can be seen in Fig.\ref{Fig1},
which shows that the sound speeds squared become
 negative at early times. This means that the early stage of the universe
  should be inhomogeneous  \cite{Starobinsky:2019xdp}.

To recapitulate, the above example shows that anisotropies in the theory with ${\cal G}=\mu+\alpha\,X$ are damped at early times. 
It is possible that choosing  other functions  ${\cal G}(X)$ yields other  models with a similar property. 
However, we shall now rather return to the original anisotropy equations \eqref{fi1} and \eqref{fi2} and consider situations 
when the nonlinear terms in these equations become important.

\section{The $G_5(X)$ case}
\setcounter{equation}{0}

Theories with a nontrivial $G_5(X)$ are also characterized by a non-constant GW speed. 
We shall consider a theory which 
also shows two inflationary stages, similarly to the $G_4(X)$ model considered above. 
It is defined by the choice
\be              \label{X}
G_2=X-\Lambda,~~~~~~G_3=0,~~~~~G_4=\frac12\,\mu,~~~~~~G_5=const+\xi\,\sqrt{2X}.
\ee
Equations \eqref{eq1},\eqref{fi1}-\eqref{fi3} then reduce to  
\bea
3\mu\,(H^2-\dot{\beta}_+^2-\dot{\beta}_-^2)+4\xi\,\dot{\phi}^2(2\dot{\beta}_+-H)[(H+\dot{\beta}_+)^2-3\dot{\beta}_-^2]&=&
\frac{1}{2}\,\dot{\phi}^2+\Lambda\,, \nn \\
\dot{\phi}\left(-1+2\xi\,(2\dot{\beta}_+-H)[(H+\dot{\beta}_+)^2-3\dot{\beta}_-^2]\right)&=&\frac{C_\phi}{{\rm a}^3}\,, \nn \\
\left(\mu\dot{\beta}_+ -\xi\,\dot{\phi}^2(\dot{\beta}_+^2+H\dot{\beta}_+-\dot{\beta}_-^2)\right)&=&\frac{C_+}{{\rm a}^3}\,, \nn \\
\left(\mu+\xi\,\dot{\phi}^2 (2\dot{\beta}_+-H))\right)\dot{\beta}_-&=&\frac{C_-}{{\rm a}^3}\,, 
\eea
all containing  terms nonlinear in $\dot\beta_\pm$. 
Their dimensionless version is obtained by 
setting 
\bea
{\rm a}={\rm a}_0\,a,~~~H=H_0\,{y},~~~~\Lambda=3\mu H_0^2\Omega_0,~~~~
\xi=-\frac{1}{8H_0^3\zeta^3},
\nn \\
C_\phi=\sqrt{-\Omega_6\mu\xi H_0}\, H_0^2 {\rm a}_0^3,~~~~C_\pm =\mu H_0 {\rm a}_0^3 Q_\pm\,,~~~~
\dot{\phi}=\sqrt{-\frac{\mu}{H_0\xi}}\,\psi\,,~~~~\dot{\beta}_\pm=H_0 s_\pm\,,
\eea
where we assume that the coupling $\xi$ is {\it negative}, hence $\zeta>0$. 
This yields the equations 
\bea                \label{eq-an}
3\,(y^2-s_+^2-s_-^2)+4\psi^2[ (y-2s_+)[(y+s_+)^2-3s_-^2]- \zeta^3]&=&3\Omega_0\,, \nn \\
\left(4\zeta^3- (y-2s_+)[(y+s_+)^2-3s_-^2]\right)\psi&=&\frac{\sqrt{\Omega_6 }}{2a^3}, \nn \\
\left(s_+ +\psi^2[s_+^2+ys_+-s_-^2  ] \right)&=&\frac{Q_+}{a^3}, \nn \\
\left(1 +\psi^2 (y-2s_+)] \right)s_-&=&\frac{Q_-}{a^3}. 
\eea
Consider first the isotropic case, 
\be
s_\pm=0,~~~~Q_\pm=0.
\ee
Then equations \eqref{eq-an} reduce to 
\bea
3y^2+4\psi^2(y^3-\zeta^3)=3\Omega_0\,, ~~~~~~
(4\zeta^3-y^3)\psi=\frac{\sqrt{\Omega_6 }}{2a^3},
\eea
with the solution 
\be                       \label{sol-is}            
a^6=\frac{\Omega_6(\zeta^3-y^3)}{3(4\zeta^3-y^3)^2(y^2-\Omega_0)},~~~~~~
\psi=\frac{\sqrt{\Omega_6} }{2a^3(4\zeta^3-y^3)}.
\ee
This solution again shows the early and late inflationary stages, since the Hubble parameter
\be
\zeta\leftarrow y=\frac{H}{H_0}\rightarrow \sqrt{\Omega_0}~~~~\text{as}~~~~~0\leftarrow a\rightarrow \infty. 
\ee
Requiring the solution to pass through the $a=y=1$ point yields 
\be               \label{Om6}
\Omega_6=\frac{3(4\zeta^3-1)^2(1-\Omega_0) }{(\zeta^3-1)}.
\ee
Choosing $\Omega_0=0.7$ and $\zeta=5$ then yields 
\begin{figure}
     \centering   
       			\resizebox{9cm}{7cm}{\includegraphics{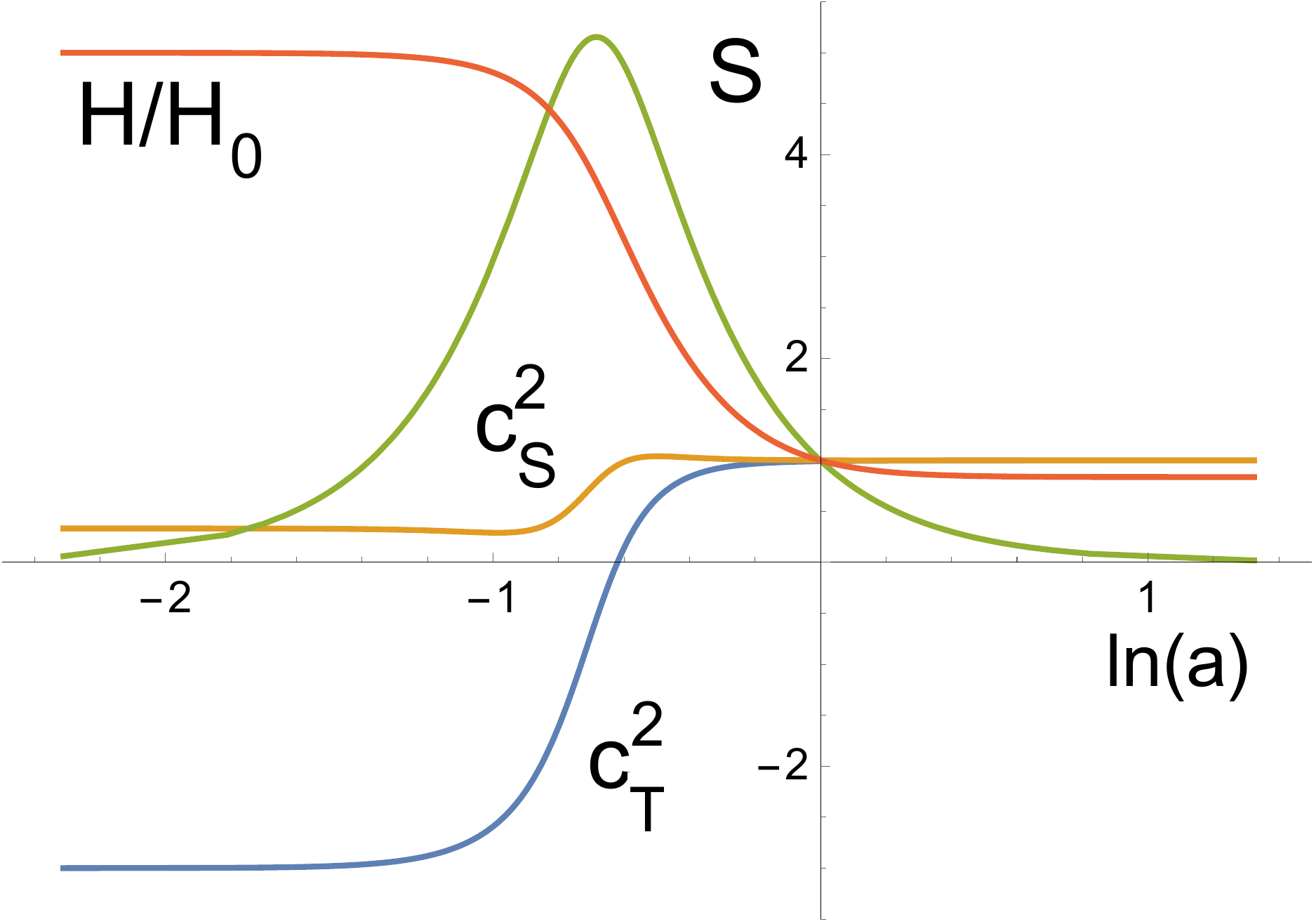}}

   \caption{The  Hubble rate, the sound speeds squared and 
   the anisotropy amplitude ${\cal S}$ defined by \eqref{sss} against $\ln(a)$ for the solution \eqref{sol-is} with 
   $\Omega_0=0.7$ and $\zeta=5$. The sound speed squared in the scalar sector is always positive.
   In the linear approximation assumed in \eqref{sss} 
  the anisotropies do not contribute to the Hubble rate. 
   }   
   \label{Fig3}
\end{figure}
the result shown in Fig.\ref{Fig3}. Remarkably, we see that the sound speed squared in the scalar sector is now always positive.
The kinetic terms are also positive, and there remains only the gradient instability in the tensor sector at early times. 
Therefore, the theory is more stable than the $G_4(X)$ model considered above. This suggests that other choices 
of functions $G_A(X)$ may perhaps give completely stable theories, but this issue requires a separate analysis.

 Equation \eqref{sol-is} 
actually defines not one by two different solutions related to each other via  $a\to -a$ and $\psi\to -\psi$, since 
$a^3=\pm\sqrt{a^6}$ can be either positive or negative whereas the metric contains only $a^2$ and  is 
insensitive to the sign of $a$. As we shall see below, the anisotropic generalizations  of these two solutions 
will no longer be related to each other in a simple way.

Let us consider anisotropic solutions of \eqref{eq-an}, starting from  the simplest case where $Q_\pm=0$. 
The simplest solution is then the isotropic one,
\be          \label{iso}
s_\pm=0,
\ee
with $a$ and $\psi$ given by \eqref{sol-is}. In addition, 
since the equations are nonlinear in the anisotropies, there are also solutions with $s_\pm\neq 0$. 
They can be represented in the parametric form, choosing  $\psi$ as the parameter:
\be               \label{aa}
a^3=\frac{\sqrt{\Omega_6}\,\psi^5 }{4\zeta^3\psi^6-3\Omega_0\psi^4+1},~~~~~
y=\frac23\,\zeta^3\psi^4+\frac12\Omega_0 \psi^2-\frac{5}{6\psi^2},
\ee
with the anisotropies being either 
\be
s_+=\frac12\left(y+\frac{1}{\psi^2} \right) ,~~~~s_-=\pm \sqrt{3}\,s_+\,,
\ee
or
\be
s_+=-\left(y+\frac{1}{\psi^2}\right),~~~~s_-=0. 
\ee
The parameter $\psi$ in  \eqref{aa} takes values in the interval $[0,\psi_{\rm m}]$ where  $\psi_{\rm m}$ is the root of 
$4\zeta^3\psi^6-3\Omega_0\psi^4+1=0$. For example, if $\zeta=0.6$ and $\Omega_0=0.7$ then $\psi_{\rm m}=0.92$. 
When $\psi$ increases from zero to   $\psi_{\rm m}$, the scale factor $a$ grows from zero to infinity, 
while the Hubble parameter $y$ and the anisotropy behave as follows: 
\be                   \label{xxx}
\infty\leftarrow y\rightarrow \frac{1-\Omega_0\psi^4_{\rm m}}{\psi^2_{\rm m}},~~~~~~~~~
-\infty\leftarrow y+\frac{1}{\psi^2} \rightarrow -\Omega_0\psi_{\rm m}^2~~~~~~~
 ~~~~\text{as}~~~~~0\leftarrow a\rightarrow \infty. 
\ee
We see that the anisotropies $s_\pm\sim (y+1/\psi^2)$ do not vanish at late times but approach constant values, 
unless for $\Omega_0=0$.  This again provides a counterexample to the standard wisdom.
Indeed, in General Relativity the Bianchi universes with a positive cosmological constant always evolve toward 
an isotropic state  at late times \cite{Starobinsky:1982mr}, \cite{Wald:1983ky}. The solution \eqref{aa}-\eqref{xxx},
although also containing a positive cosmological constant, shows just the opposite ``self-anisotropizing" behaviour. 
It should be said  that such a self-anisotropization  in the Horndeski theory 
with a non-trivial $G_5(X)$ was actually detected before in Ref. \cite{Tahara:2018orv}, also when analyzing 
the  Bianchi I models. We therefore shall not discuss  this phenomenon anymore and simply refer to \cite{Tahara:2018orv}, 
since we are interested in the 
early time ``isotropization" rather than  in the late time ``anisotropization". 
For all other solutions that we consider in this text, apart from \eqref{aa}-\eqref{xxx}, the anisotropies always approach zero at late times. 
Therefore,  we now return back to the isotropic solution \eqref{iso} and consider 
its  deformations induced by adding nonzero  anisotropy charges $Q_\pm$. 

\begin{figure}
     \centering   
       			\resizebox{8cm}{6cm}{\includegraphics{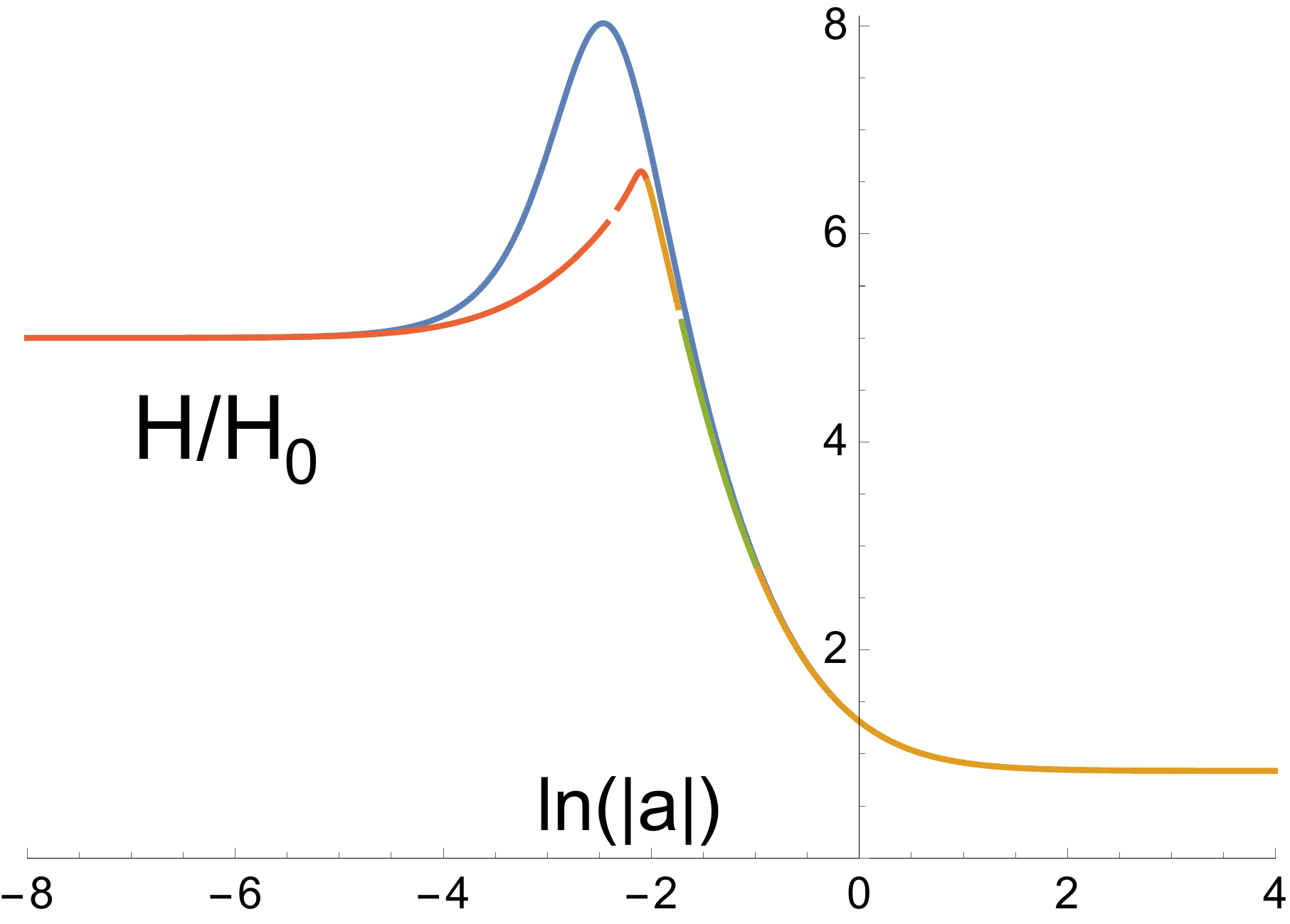}}

   \caption{The Hubble rate $H/H_0=y=Y-s_+$  against $\ln(|a|)$ 
   for the two solutions
   of \eqref{last}  with 
   $\Omega_0=0.7$, $\zeta=5$ for  $S=0.02$. The Hubble rate is affected by the anisotropies when  the nonlinear
   terms are taken into account.}   
   \label{Fig4}
\end{figure}

If $Q_\pm$ are very small, then one can expect the anisotropies $s_\pm$ to be small as well, in which case one can neglect 
all nonlinear in $s_\pm$  terms in the equations. The first two equations in \eqref{eq-an} contain only such terms,
and neglecting them yields the equations of the isotropic case whose solution was described above by \eqref{sol-is}. 
The last two equations in \eqref{eq-an} do contain terms linear in  $s_\pm$, and keeping only these yields the solution  
\be           \label{sss}
s_\pm=\frac{Q_\pm}{a^3(1+\psi^2 y)}\equiv Q_\pm {\cal S}, 
\ee
with  $a,\psi$ are given by \eqref{sol-is}. The function ${\cal S}$ here is well localized, as seen in Fig.\ref{Fig3},
and it has the following limits:
\be
\frac{36\zeta^5}{\Omega_6}\, a^3\leftarrow {\cal S} \rightarrow \frac{1}{a^3}
 ~~~~\text{as}~~~~~0\leftarrow a\rightarrow \infty. 
\ee
Therefore, the anisotropies are suppressed both at early and late times.

If the charges $Q_\pm$ are not  small, then one cannot neglect the nonlinear in anisotropies terms in the equations. 
It is not then obvious that the anisotropies will still be suppressed  at early and late times. 
Let us therefore take  the nonlinear terms into account. 
To simplify the analysis, we set one of the anisotropy amplitudes and the corresponding charge to zero, 
\be
s_-=Q_-=0,
\ee
while keeping $s_+\neq 0$ and denoting 
\be
Q_+=\sqrt{\Omega_6}\,S.
\ee
It turns out that all nonlinear in $s_+$ terms in the equations  can be absorbed by introducing the new variable 
\be
Y=y+s_{+}.
\ee
Then equations \eqref{eq-an} reduce, without any approximation,  to
\bea
4\psi^2( Y^3-3Y^2s_+-\zeta^3)+3Y^2-6Ys_+&=&3\Omega_0\,,  \nn \\
\psi(4\zeta^3+3Y^2s_+-Y^3)&=&\frac{\sqrt{\Omega_6}}{2a^3}\,,  \nn \\
(Y\psi^2+1)s_+&=&\frac{S\sqrt{\Omega_6}}{a^3}. 
\eea
Their solution is 
\bea           \label{last0}
a^3=\frac{\sqrt{\Omega_6}\,(6S\,Y^2\psi-Y\psi^2-1 ) }{2\psi(Y^3-4\zeta^3)(Y\psi^2+1) },~~~~~~~~
s_+=\frac{S\sqrt{\Omega_6}}{a^3(Y\psi^2+1)}\,,
\eea
where $Y$ and $\psi$ are related via 
\be             \label{last}
\psi^2=\frac34\,\frac{Y^2-\Omega_0}{\zeta^3-Y^3}
+\frac{3}{2}\,S\,Y\psi\frac{12\zeta^3 Y\psi^2+Y^3+8\zeta^3-3\Omega_0 Y }{(Y\psi^2+1)(Y^3-\zeta^3)}.
\ee
If $S=0$ then $s_+=0$, $Y=y$, and   these formulas reduce to \eqref{sol-is} describing 
the isotropic solution with $y=Y\in[\sqrt{\Omega_0},\zeta]$. 
If $S\neq 0$ then \eqref{last} yields  the fourth order algebraic equation for $\psi=\psi(Y)$.  
Fortunately, all of its four solutions 
can be found 
analytically. 
Each of them
is real-valued only within a finite interval of $Y$, but  combining these piecewise solutions together yields 
two global solutions which are smooth and real-valued everywhere in the interval $Y\in[\sqrt{\Omega_0},\zeta]$. 
These two solutions have opposite signs of $\psi$ and of $a$. 

\begin{figure}
     \centering   
       			\resizebox{8cm}{7cm}{\includegraphics{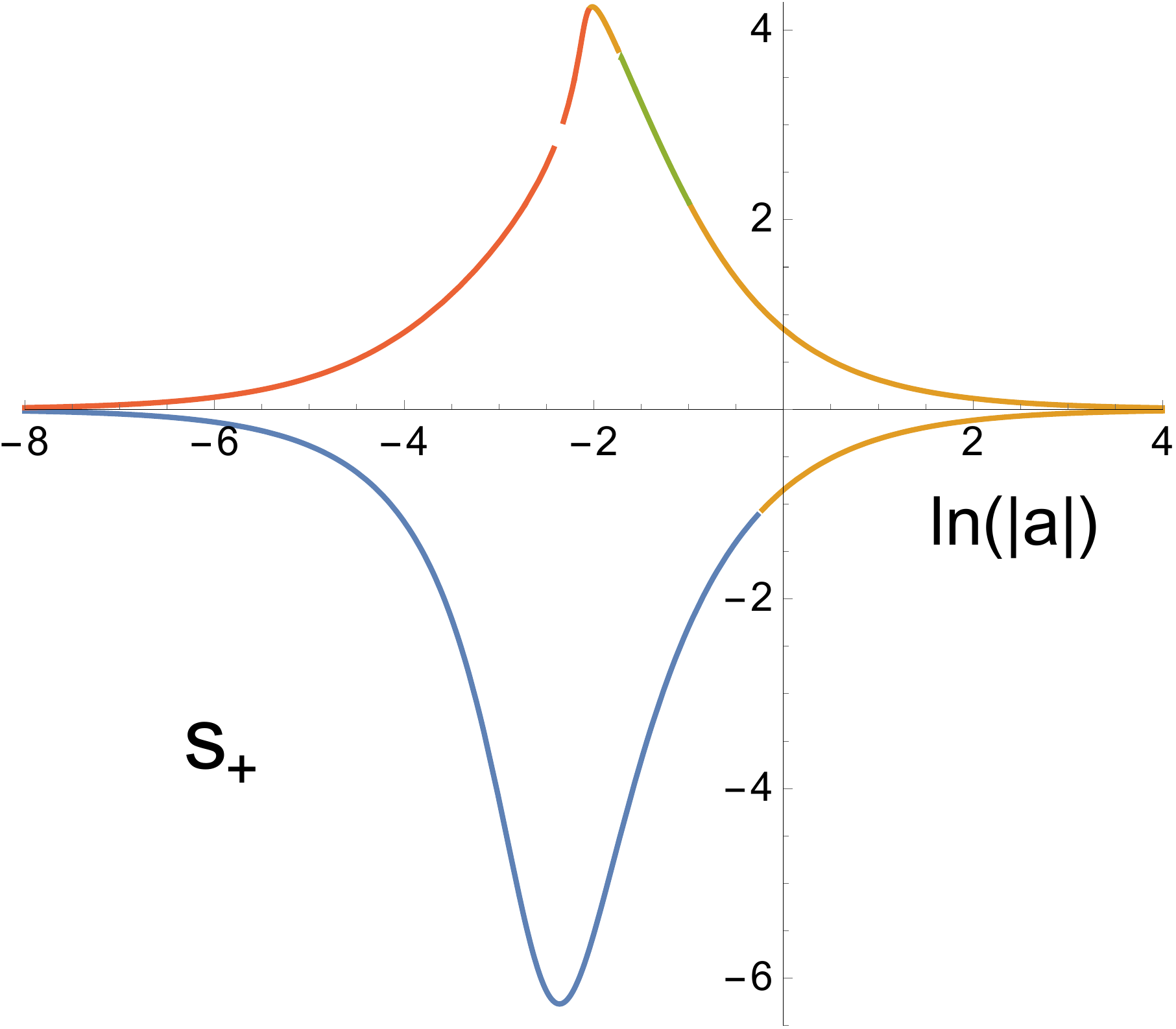}}
			
   \caption{The anisotropy $s_+$ defined by \eqref{last0} against $\ln(|a|)$ for the two solutions shown in Fig.\ref{Fig4},
   assuming the normalizaton \eqref{Om6}.}   
   \label{Fig5}
\end{figure}

In the isotropic limit these two solutions are related by simply $\psi\to -\psi$ and $a\to -a$, as described above,  while 
 their Hubble rates $y(|a|)$ are the same.
If $S\neq 0$ then the two solutions are no longer related to each other in a simple way 
and their Hubble rates are different, as seen in  Figs.\ref{Fig4}. 
As seen in Fig.\ref{Fig5}, the anisotropies again vanish at late and early times. 
These nonlinear solutions were obtained for the value of the anisotropy parameter which is still small enough, 
$S=0.02$, but increasing $S$ does not qualitatively change the situation. 
Aalready for $S=1$  the anisotropy $s_+$ attains very large values in the intermediate region,
but it always approaches zero as $a\to 0,\infty$. Therefore, the anisotropies are damped at early times 
also at the nonlinear level.

\section{Conclusions}
\setcounter{equation}{0}

Summarizing the above discussion, we have studied homogeneous and anisotropic Bianchi~I  cosmologies within  the
most general Horndeski class. Our aim was to see whether the phenomenon of anisotropy damping 
previously observed  within  the  specific Horndeski model \cite{Starobinsky:2019xdp} is present in other Horndeski
theories  as well. 
We have found the phenomenon  to be  absent for a large class of Horndeski models in which the GW speed is constant. 
However, the phenomenon  seems to be generically present in the more general models with nontrivial $G_4(X)$ and/or $G_5(X)$. 
The GW speed in such theories is not constant, but no contradiction with the observation arises since 
the predicted value of the GW speed {\it at present} is extremely close to unity, whereas  no observation data of the GW speed
in the past are available. 

Such theories  show gradient instabilities at early times, therefore their initial  phase, although not anisotropic,
cannot be isotropic either. It should therefore be inhomogeneous. At the same time, it is possible that a systematic 
analysis of theories with more general $G_4(X,\phi)$ and/or $G_5(X,\phi)$ may reveal models free of instabilities. 
In the case of nonsingular  bounce-type \cite{Battefeld:2014uga} or Genesis-type 
\cite{Creminelli:2010ba}) cosmologies, 
no stable solution can exist within the Horndeski class \cite{Libanov:2016kfc}, \cite{Kobayashi:2016xpl}, 
 although  they exist within the more general DHOST models (see \cite{Mironov:2019haz} for a review). 
 However, we are unaware of similar no-go results for cosmologies with an initial singularity. 
 In fact, an explicit example of a completely stable Horndeski theory is known,
 although not containing an early inflationary phase 
 \cite{Koutsoumbas:2017fxp}. Therefore, it is not excluded 
 that stable cosmologies with the  early and late inflationary phases may exist within the Horndeski theory, hence   their 
 anisotropies should be damped near singularity.   
 
It should also be mentioned that, as was first observed in \cite{Starobinsky:2019xdp}, the effect of anisotropy damping  
may be sensitive to the inclusion of spatial curvature.

\begin{acknowledgements}
It is a pleasure to thank Karim Noui for discussions. 
R.G., R.M., A.A.S. and S.V.S were supported by the Russian Foundation for Basic Research, grant No.19-52-15008.
M.S.V. was partly supported by the  CNRS/RFBR PRC grant No.289860. 
This work was  also partially  supported by the Kazan Federal University 
Strategic Academic Leadership Program. 
\end{acknowledgements}


\providecommand{\href}[2]{#2}\begingroup\raggedright\endgroup

\end{document}